\documentclass{article}

\usepackage[main, final]{neurips_2025}

\usepackage[utf8]{inputenc} 
\usepackage[T1]{fontenc}    
\usepackage{hyperref}       
\usepackage{url}            
\usepackage{booktabs}       
\usepackage{amsfonts}       
\usepackage{nicefrac}       
\usepackage{microtype}      
\usepackage{xcolor}         
\usepackage{graphicx}
\usepackage{subfig}

\title{State Space Models for Bioacoustics: A Comparative Evaluation with Transformers}

\author{%
  Chengyu Tang\thanks{Corresponding author.}\\
  Auburn University\\
  Auburn, AL 36849 \\
  \texttt{tangcy@auburn.edu} \\
  \And
  Sanjeev Baskiyar\\
  Auburn University\\
  Auburn, AL 36849 \\
  \texttt{baskisa@auburn.edu}
}

\begin{document}

\maketitle

\begin{abstract}

In this study, we evaluate the efficacy of the Mamba architecture bioacoustics by introducing BioMamba, a Mamba-based audio representation model for wildlife sounds. We pre-train a BioMamba using self-supervised learning on a large audio corpus and evaluate it on the BEANS benchmark across diverse classification and detection tasks. Compared to the state-of-the-art Transformer-based model (AVES), BioMamba achieves comparable performance while significantly reducing VRAM consumption. Our results demonstrate Mamba's potential as a computationally efficient alternative for real-world environmental monitoring.

\end{abstract}

\section{Motivation}

In recent decades, deep learning has emerged as a powerful methodology, finding widespread applications across diverse domains and data modalities. Typically, deep learning models are trained using large amounts of annotated data. However, acquiring high-quality annotated data remains challenging, particularly due to the extensive resources and human effort required for labeling. Transformer-based models, such as BERT \cite{devlin2019bert}, ViT \cite{dosovitskiy2021an}, have achieved remarkable progress through self-supervised pre-training. Nevertheless, Transformer models inherently possess quadratic computational complexity due to the self-attention mechanism

In bioacoustics, labeled data is particularly limited. Bioacoustic data poses unique challenges, including the need for computationally efficient models capable of handling long-term audio sequences effectively. While some studies, such as AVES \cite{hagiwara2023aves}, have begun to address bioacoustic modeling, the high computational demands of traditional Transformer models significantly limit their practical utility in real-world bioacoustic scenarios. Although there have been various attempts to lower the complexity of self-attention and increase its efficiency, such as the Reformer~\cite{kitaev2020reformer}, Linformer~\cite{wang2020linformer}, and Flash Attention~\cite{dao2022flashattention}, they still do not fully overcome the fundamental efficiency limitations inherent to self-attention mechanisms.

One of the latest attempts is the Mamba model~\cite{gu2023mamba}, which is based on the State Space Model (SSM). Mamba is able to achieve Transformer-level performance in various domains, including audio representation learning \cite{shams2024ssamba,jiang2025speech,lieber2024jamba} while significantly improving computational efficiency with its linear complexity. Preliminary studies are exploring Mamba for bioacoustic applications that have shown encouraging signs in bioacoustics \cite{mu2024seld}, but they remain limited and small-scale.

Motivated by these observations and gaps, this study investigates the effectiveness of a Mamba-based model for bioacoustic tasks. We trained a Mamba-based model, BioMamba, using self-supervised learning for animal sound modeling and evaluated it on a comprehensive bioacoustic benchmark, including animal sound classification and detection. We compared its performance with multiple baseline models, from traditional machine learning models to a state-of-the-art Transformer-based model. To our best knowledge, BioMamba is the first Mamba-based model trained on large-scale audio datasets and comprehensively evaluated across multiple bioacoustic tasks. The results demonstrate that the Mamba-based model can achieve comparable performance to a Transformer-based model while consuming significantly less memory, which implies its potential in advancing bioacoustic research and real-world environmental monitoring applications.

\section{Related work}

\subsection{Transformer-based audio representation models}

Recent years have seen a surge of transformer-based models for audio representation learning, inspired by the success of transformers in NLP and vision. In speech processing, pioneering self-supervised models like wav2vec 2.0 \cite{baevski2020wav2vec2} and HuBERT \cite{hsu2021hubert} leverage unlabeled audio at scale to learn powerful feature encoders. Wav2Vec 2.0 introduced a contrastive learning framework on masked speech audio and demonstrated that pre-training on tens of thousands of hours of unlabeled speech enabled automatic speech recognition (ASR) with very little labeled data. Building on this, HuBERT uses masked prediction of latent acoustic units, where audio frames are clustered and a transformer model is trained to predict these cluster labels for masked timesteps. HuBERT achieved state-of-the-art performance on the SUPERB benchmark for speech tasks, demonstrating the efficacy of learning rich acoustic representations from self-supervision. Beyond speech, similar transformer architectures have been applied to general audio. For example, the Self-Supervised Audio Spectrogram Transformer (SSAST) \cite{gong2022ssast} extends masked spectrogram modeling to AudioSet, learning transferable features for a variety of sound classification tasks. Overall, transformer audio models pre-trained on large-scale audio data have become a standard approach for audio-related tasks with limited labeled data.

In the bioacoustics domain, collecting extensive labeled data for animal vocalizations is challenging, which makes self-supervised learning especially appealing. Researchers have thus begun to apply audio foundation models, such as HuBERT, to bioacoustic tasks, including species sound classification and call detection. Hagiwara introduced AVES \cite{hagiwara2023aves}, a transformer-based audio model specifically tailored to wildlife sounds. AVES was pre-trained on a diverse collection of unannotated animal audio (along with some speech and environmental audio) using a HuBERT-style masked prediction objective. Fine-tuned on downstream bioacoustic tasks, AVES significantly outperformed supervised CNN models on the BEANS benchmark \cite{hagiwara2023beans}, a suite of bird, frog, and marine mammal sound classification and detection datasets. These results demonstrate that self-supervised representations originally developed for speech can be transferred effectively to bioacoustics, thereby closing the gap in data-scarce regimes. 

\subsection{Mamba-based audio models}

While transformers have dominated sequence modeling, a new class of architectures based on state-space models has emerged as an efficient alternative for long audio sequences. In particular, the Mamba architecture \cite{gu2023mamba} integrates structured state-space models (SSM) with a time-varying mechanism, achieving linear time sequence modeling without self-attention. Mamba was first shown to match or exceed Transformers on long-sequence tasks in vision and text, and has since been explored in speech and audio applications. For example, Zhang et al. \cite{zhang2024mamba} replaced the transformer encoder with Mamba in ASR and speech processing pipelines, and found that Mamba-based models can excel in reconstruction tasks, such as speech enhancement or masked spectrogram recovery, but may require additional decoding modules to meet or beat transformers on classification tasks like speech recognition.

Importantly, Mamba-based models have been developed for self-supervised audio representation learning. SSAMBA \cite{shams2024ssamba} is a Mamba-based encoder trained on masked spectrogram prediction using the AudioSet corpus. It outperformed a comparable transformer (SSAST) by a large margin on a suite of ten audio tasks. It also proved more data-efficient, achieving stronger results than the transformer when both were trained on smaller subsets of data. These findings imply that Mamba-based state-space models are promising alternatives to transformers in audio-related areas. Concurrently, other studies have applied Mamba to specialized speech problems. For instance, Speech Slytherin \cite{jiang2025speech} evaluated Mamba’s performance on speech separation, recognition, and synthesis tasks, reporting favorable speed-ups and solid accuracy compared to attention-based baselines. Likewise, Mamba has been integrated into speech enhancement models and even hybrid Transformer-Mamba architectures \cite{lieber2024jamba} to combine the strengths of both paradigms. In summary, the Mamba architecture has rapidly gained traction in audio research as an efficient sequence model, and early results show it can achieve high performance on par with or surpassing transformers across a range of speech and sound tasks.

The application of Mamba-based models to bioacoustics is only just beginning to be explored. There are limited preliminary studies that have shown encouraging signs in bioacoustics. One relevant example is SELD-Mamba \cite{mu2024seld}, a Mamba-based network for sound event localization and detection (SELD). In this model, the standard Conformer blocks of a SELD system were replaced with bidirectional Mamba blocks to capture long-range context efficiently. Using a two-stage training scheme, SELD-Mamba achieved strong results on the DCASE 2024 challenge dataset, matching the accuracy of transformer-based SELD while reducing computational complexity. This demonstrates the potential of state-space models to handle acoustic event detection tasks that are conceptually similar to bioacoustic call detection (e.g., detecting sound events in continuous audio streams).

To the best of our knowledge, there are no published studies that specifically deploy a Mamba architecture for animal sound classification or detection. Existing models are mostly evaluated on small datasets, which makes this an open research direction.

\section{Methodology}

\subsection{Model architecture}

The BioMamba model consists of two main components: a feature extractor and an encoder. The feature extractor is responsible for converting the raw audio waveform into high-level features, which are then processed by the subsequent encoder. 

\paragraph{Feature Extractor}

To process the raw data, we first need to convert it into tokens that the model can process. This is done by extracting features from the raw audio waveform using a convolutional neural network (CNN)-based feature extractor, which is a common practice in the literature. Specifically, we employ the same feature extractor architecture as used in the wav2vec 2.0 model \cite{baevski2020wav2vec2}. It consists of 7 one-dimensional convolution layers and has a fixed down-sampling rate of 320. Therefore, the input audio clips must have the same sample rate to maintain performance. In this study, we use 16 kHz as the target sample rate, a common practice in the literature. As a result, the extracted feature is a sequence of frames, each corresponding to a 20-ms segment of the original audio.

\paragraph{Encoder}

Following the CNN-based feature extractor, the resulting sequence of feature vectors is processed by a series of stacked Mamba \cite{gu2023mamba} layers to model the long-range temporal dependencies in the audio. The Mamba architecture is based on State Space Models(SSMs) and is designed to efficiently handle long sequences with linear complexity in both memory and computation. Each Mamba layer processes input sequences by maintaining a hidden state that evolves through linear recurrent transitions combined with efficient gating mechanisms, enabling effective modeling of both local and long-range dependencies in sequential data. Specifically, Mamba utilizes selective gating and convolutional state-space modeling techniques to adaptively control information flow and capture temporal dependencies at different scales.

In this study, we adopt the latest Mamba2 \cite{dao2024mamba2} architecture and stack multiple Mamba layers sequentially to form a deep model, which allows the model to progressively refine feature representations.

\subsection{Training}
\paragraph{Pre-training}

We employ a similar pre-training scheme as HuBERT \cite{hsu2021hubert} and AVES \cite{hagiwara2023aves} using self-supervised learning. Given the extracted frame sequence from the CNN feature extractor, a subset of the frames is randomly masked, and the model's goal is to predict the masked frames. Unlike NLP, a pre-defined vocabulary is unavailable. Instead, we apply K-means clustering to the frames and use their cluster IDs as pseudo-labels. The model is then trained to predict the pseudo-labels of the masked frames. This approach allows the model to learn meaningful representations of audio data without requiring explicit labels.

To update the model, we apply gradient descent on the cross-entropy loss between the prediction and the actual pseudo-labels over all the masked frames.

\[ L_m = \sum_{t \in M} \mathrm{log}p_f(z_t|\tilde{X},t) \]

\noindent where $M$ is the set of indices of the masked frames and $\tilde{X}$ is the masked version of the input sequence. 

Depending on the characteristics we utilize to cluster the audio frames, the pre-training process is divided into two phases. 

During the first phase of pre-training, the pseudo-labels are generated by applying k-means clustering on the MFCCs of the audio data. The MFCCs are a widely used compact feature representation in audio processing. In this study, the MFCCs are computed for every 25 ms window with a 10 ms overlap, resulting in a downsampling factor of 160.

After the first pre-training phase, the model has learnt to extract meaningful features from the audio data, which means that we can utilize the high-level features extracted by the model itself to generate more reliable pseudo-labels that incorporate more contextual information instead of relying solely on the MFCCs. Therefore, in the second iteration, we apply k-means to the internal representations of the 6th Mamba layer of the model. 

\paragraph{Fine-tuning}

In this study, the pre-trained model is fine-tuned for downstream bioacoustic classification and detection tasks using supervised learning, where each audio clip is associated with a label. For classification, each label refers to the species or individual that produces the sound. For detection, each label is a binary list, indicating whether each class is present in the corresponding input sound. To accommodate the downstream tasks, the pre-trained model is extended with a task-specific single-layer classification head. Since the encoding output of the pretrained model is a matrix of shape $[T, d]$, where $T$ is the sequence length or number of frames, and $d$ is the model's internal dimension, the encoding is mean-pooled along the sequence dimension before being passed to the classification head. We use cross-entropy loss for classification and binary cross-entropy loss (averaged over all classes) for detection, and gradient descent to optimize the model. Like previous studies, the weights of the CNN feature extractor are not updated during the fine-tuning process.

\section{Evaluation} \label{sec:evaluation}

\subsection{Training details} \label{sec:training_details}

The BioMamba model is implemented based on the official Mamba repository \footnote{\url{https://github.com/state-spaces/mamba}} and TorchAudio's example training scripts \footnote{\url{https://github.com/pytorch/audio/tree/main/examples/hubert}}. For AVES, multiple pre-trained models with different sizes and training data are provided by the author \footnote{\url{https://github.com/earthspecies/aves}}. We selected the \texttt{bio} version, which the author claims has the best overall performance \cite{hagiwara2023aves}.

Similar to AVES, the model is trained for 100,000 steps during both stages of the pre-training, using a batch size of 700 seconds of audio. We used a peak learning rate of $5 \times 10^{-4}$, with a linear warmup for the initial 5\% steps and a linear decay for the remaining steps.

For fine-tuning, we used the data splits pre-defined in each sub-dataset. We swept the learning rate over $1.0 \times 10^{-5}$, $5.0 \times 10^{-5}$, and $1.0 \times 10^{-4}$, each for 50 epochs and selected the model with the best validation metric.

\subsection{Datasets}

To pre-train BioMamba, we use a collection of three datasets, FSD50K \cite{fonseca2022fsd50k}, VGGSound \cite{chen2020vggsound}, and AudioSet \cite{gemmeke2017audioset}. For VGGSound and AudioSet, we use their subsets corresponding to animal sounds.

To evaluate the generalization capability of the pre-trained BioMamba model, we conduct a comprehensive evaluation on different tasks and datasets. The platform we use is the BEANS benchmark \cite{hagiwara2023beans}. It provides a comprehensive evaluation of animal sound classification and detection tasks, curating 12 datasets in various species, including birds, frogs, and marine mammals. For classification, the model is trained to identify the species or individual of the emitter given an audio input. For detection, the model is trained to determine whether each species or individual is present in the given audio clip.

Datasets for classification include: \texttt{bats} \cite{prat2017annotated}, \texttt{cbi} \cite{cornell2020cbi}, \texttt{dogs} \cite{yin2004barking}, \texttt{humbug} \cite{kiskin2021humbugdb}, and \texttt{watkins} \cite{sayigh2016watkins}. Datasets for detection include: \texttt{dcase} \cite{morfi2021few}, \texttt{enabirds} \cite{chronister2021annotated}, \texttt{gibbons} \cite{dufourq2021automated}, \texttt{hiceas} \cite{noaa2022hawaiian}, and \texttt{rfcx} \cite{lebien2020pipeline}..

\subsection{Main results}

We compare BioMamba with a wide range of baseline models: XGBoost, ResNet \cite{he2016deep} (18, 50, and 152 layers), VGGish \cite{hershey2017cnn}, and AVES \cite{hagiwara2023aves}.

The detailed comparison is shown in Table~\ref{tab:results} \footnote{Class-based mAP is typically calculated using the predicted probabilities, while in the official BEANS repository, the mAP of machine learning models is calculated using their binary predictions. Our results are consistently calculated based on the conventional mAP formula. As a result, some values are noticeably higher than those reported in the original paper. For more details, refer to \url{https://github.com/earthspecies/beans/issues/3}}. Overall, BioMamba achieves comparable performance with AVES and outperforms other models in most cases. It is worth noting that the hyperparameters for BioMamba are determined primarily to ensure consistency with the baseline model. With more thorough hyperparameter tuning, we believe its performance can be further improved.

\begin{table*}[t]
    \caption{Evaluation Results on the classification tasks in the BEANS benchmark. We use accuracy for classification and mean average precision (mAP) for detection as the evaluation metric.}
    \label{tab:results}
    \centering
    \begin{tabular}{l|ccccc|ccccc}
        \hline
         & \multicolumn{5}{c|}{\textbf{Classification}} & \multicolumn{5}{c}{\textbf{Detection}} \\
         \hline
         & \texttt{bats} & \texttt{cbi} & \texttt{dogs} & \texttt{humbug} & \texttt{wtkns} & \texttt{dcase} & \texttt{enab} & \texttt{gib} & \texttt{hiceas} & \texttt{rfcx} \\
        \hline
        LR       & 0.661 & 0.159 & 0.892 & 0.751 & 0.770 & 0.217 & 0.343 & 0.012 & 0.345 & 0.084 \\
        SVM      & 0.720 & 0.136 & \underline{0.914} & 0.779 & 0.870 & 0.231 & 0.430 & 0.082 & 0.461 & 0.093 \\
        DT       & 0.477 & 0.024 & 0.626 & 0.697 & 0.643 & 0.120 & 0.221 & 0.010 & 0.219 & 0.014 \\
        GBDT     & 0.682 & 0.040 & 0.820 & 0.758 & 0.779 & 0.144 & 0.331 & 0.010 & 0.381 & 0.024 \\
        XGBoost  & 0.700 & 0.100 & 0.820 & 0.771 & 0.811 & 0.200 & 0.394 & 0.010 & 0.366 & 0.061 \\
        RN18     & 0.468 & 0.287 & 0.640 & 0.707 & 0.755 & 0.145 & 0.312 & 0.232 & 0.263 & 0.072 \\
        RN50     & 0.446 & 0.325 & 0.547 & 0.682 & 0.732 & 0.145 & 0.271 & 0.176 & 0.218 & 0.047 \\
        RN152    & 0.474 & 0.351 & 0.468 & 0.667 & 0.711 & 0.151 & 0.292 & 0.192 & 0.227 & 0.065 \\
        VGGish   & \textbf{0.750} & 0.447 & 0.906 & \underline{0.798} & 0.855 & 0.372 & 0.531 & 0.112 & 0.500 & \textbf{0.142} \\
        \hline
        AVES     & \underline{0.739} & \textbf{0.578} & \textbf{0.942} & \textbf{0.812} & \underline{0.879} & \underline{0.409} & \underline{0.538} & \underline{0.304} & \textbf{0.683} & \underline{0.134} \\
        \hline
        BioMamba & 0.725 & \underline{0.513} & \underline{0.914} & 0.770 & \textbf{0.896} & \textbf{0.426} & \textbf{0.548} & \textbf{0.318} & \underline{0.643} & \textbf{0.142} \\
        \hline
    \end{tabular}
\end{table*}

\subsection{Memory efficiency comparison} \label{sec:vram_comparison}

\begin{figure}[t]
    \centering
    \subfloat[Peak allocated VRAM]{
        \includegraphics[width=0.4\textwidth]{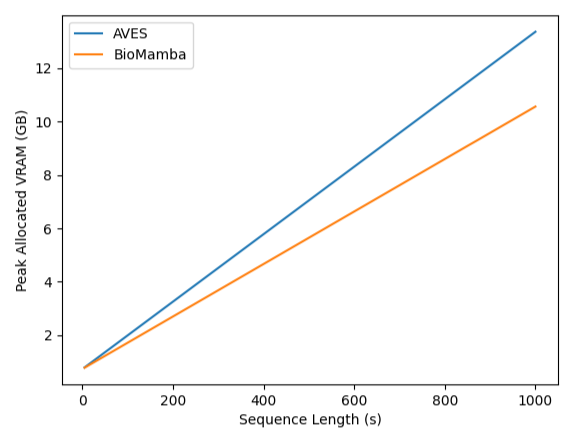}
        \label{fig:allocated_vram}
    }
    \subfloat[Peak reserved VRAM]{
        \includegraphics[width=0.4\textwidth]{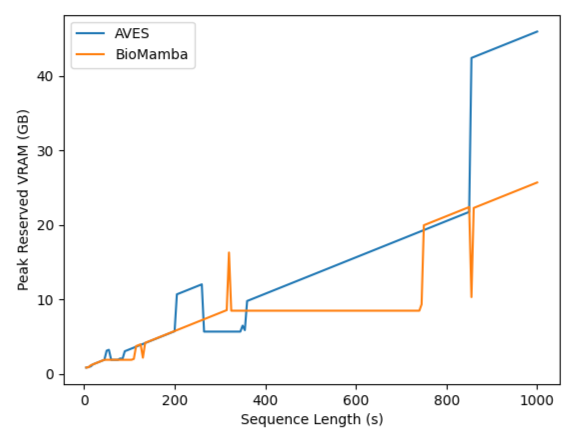}
        \label{fig:reserved_vram}
    }
    \caption{Memory usage comparison between AVES and BioMamba}
    \label{fig:vram_usage}
\end{figure}

In addition to accuracy and mean average precision, we also evaluated the inference memory efficiency of BioMamba and compared it with the Transformer-based AVES model. We tested input sequences with various lengths, ranging from 5 to 1000 seconds, with a step size of 5 seconds. For each length, we created a synthetic sequence with a 16kHz sample rate and recorded the peak reserved and allocated VRAM by the Transformer layers and Mamba layers, respectively. The experiment was conducted on a Linux server equipped with an Nvidia H100 GPU, so FlashAttention and memory-efficient kernels are enabled for AVES. From Figure~\ref{fig:vram_usage}, it can be observed that both models' memory usage grows near linearly as the sequence length increases. Even with the optimizations, AVES consistently consumes around 40\% more VRAM than BioMamba.

\section{Conclusions} \label{sec:conclusion}

In this paper, we study the effectiveness of the Mamba model in Bioacoustic applications with its efficient linear complexity. We pre-trained BioMamba using self-supervised learning and evaluated it using supervised learning on the BEANS benchmark. The results show that Mamba is competitive with Transformer in Bioacoustic tasks while using significantly less memory, implying its potential in this field.

\bibliographystyle{plain}
\bibliography{references}

\end{document}